\documentclass[useAMS,usenatbib]{mn2e}
\usepackage{tabularx}
\usepackage{xspace}
\usepackage{graphicx}
\newcommand{\ignore}[1]{}
\providecommand{\ao}{}
\renewcommand{\ao}{adaptive optics (AO)\renewcommand{\ao}{AO\xspace}\renewcommand{\Ao}{AO\xspace}\xspace}
\newcommand{\Ao}{Adaptive optics (AO)\renewcommand{\ao}{AO\xspace}\renewcommand{\Ao}{AO\xspace}\xspace}
\newcommand{\wfs}{wavefront sensor (WFS)\renewcommand{\wfs}{WFS\xspace}\renewcommand{\wfss}{WFSs\xspace}\xspace}
\newcommand{\wfss}{wavefront sensors (WFSs)\renewcommand{\wfs}{WFS\xspace}\renewcommand{\wfss}{WFSs\xspace}\xspace}
\newcommand{\shwfs}{Shack-Hartmann \wfs (SHWFS)\renewcommand{\shwfs}{SHWFS\xspace}\xspace}
\newcommand{\dm}{deformable mirror (DM)\renewcommand{\dm}{DM\xspace}\renewcommand{\dms}{DMs\xspace}\renewcommand{\Dms}{DMs\xspace}\renewcommand{\Dm}{DM\xspace}\xspace}
\newcommand{\dms}{deformable mirrors (DMs)\renewcommand{\dm}{DM\xspace}\renewcommand{\dms}{DMs\xspace}\renewcommand{\Dms}{DMs\xspace}\renewcommand{\Dm}{DM\xspace}\xspace}
\newcommand{\Dms}{Deformable mirrors (DMs)\renewcommand{\dm}{DM\xspace}\renewcommand{\dms}{DMs\xspace}\renewcommand{\Dms}{DMs\xspace}\renewcommand{\Dm}{DM\xspace}\xspace}
\newcommand{\Dm}{Deformable mirror (DM)\renewcommand{\dm}{DM\xspace}\renewcommand{\dms}{DMs\xspace}\renewcommand{\Dms}{DMs\xspace}\renewcommand{\Dm}{DM\xspace}\xspace}

\newcommand{\shs}{Shack-Hartmann sensor (SHS)\renewcommand{\shs}{SHS\xspace}\renewcommand{\shss}{SHSs\xspace}\xspace}
\newcommand{\shss}{Shack-Hartmann sensors (SHSs)\renewcommand{\shs}{SHS\xspace}\renewcommand{\shss}{SHSs\xspace}\xspace}
\newcommand{\lgs}{laser guide star
  (LGS)\renewcommand{\lgs}{LGS\xspace}\renewcommand{\lgss}{LGSs\xspace}\xspace}
\newcommand{\lgss}{laser guide stars (LGSs)\renewcommand{\lgs}{LGS\xspace}\renewcommand{\lgss}{LGSs\xspace}\xspace}
\newcommand{\Ngs}{Natural guide star (NGS)\renewcommand{\ngs}{NGS\xspace}\renewcommand{\Ngs}{NGS\xspace}\renewcommand{\ngss}{NGSs\xspace}\xspace}
\newcommand{\ngs}{natural guide star (NGS)\renewcommand{\ngs}{NGS\xspace}\renewcommand{\Ngs}{NGS\xspace}\renewcommand{\ngss}{NGSs\xspace}\xspace}
\newcommand{\ngss}{natural guide stars (NGSs)\renewcommand{\ngs}{NGS\xspace}\renewcommand{\Ngs}{NGS\xspace}\renewcommand{\ngss}{NGSs\xspace}\xspace}
\newcommand{\mems}{Micro-Electro-Mechanical Systems (MEMS)\renewcommand{\mems}{MEMS\xspace}\xspace}
\newcommand{\snr}{signal to noise ratio (SNR)\renewcommand{\snr}{SNR\xspace}\xspace}
\newcommand{\Moao}{Multi-object \ao (MOAO)\renewcommand{\moao}{MOAO\xspace}\renewcommand{\Moao}{MOAO\xspace}\xspace}
\newcommand{\moao}{multi-object \ao (MOAO)\renewcommand{\moao}{MOAO\xspace}\renewcommand{\Moao}{MOAO\xspace}\xspace}
\newcommand{\mcao}{multi-conjugate adaptive optics (MCAO)\renewcommand{\mcao}{MCAO\xspace}\xspace}
\newcommand{\ltao}{laser tomographic adaptive optics (LTAO)\renewcommand{\ltao}{LTAO\xspace}\xspace}
\newcommand{\cpu}{central processing unit (CPU)\renewcommand{\cpu}{CPU\xspace}\renewcommand{\cpus}{CPUs\xspace}\xspace}
\newcommand{\cpus}{central processing units (CPUs)\renewcommand{\cpu}{CPU\xspace}\renewcommand{\cpus}{CPUs\xspace}\xspace}
\newcommand{\psf}{point spread function (PSF)\renewcommand{\psf}{PSF\xspace}\renewcommand{\psfs}{PSFs\xspace}\renewcommand{\Psf}{PSF\xspace}\xspace}
\newcommand{\psfs}{point spread functions (PSFs)\renewcommand{\psf}{PSF\xspace}\renewcommand{\psfs}{PSFs\xspace}\renewcommand{\Psf}{PSF\xspace}\xspace}
\newcommand{\Psf}{Point spread function (PSF)\renewcommand{\psf}{PSF\xspace}\renewcommand{\psfs}{PSFs\xspace}\renewcommand{\Psf}{PSF\xspace}\xspace}
\newcommand{\fpga}{field programmable gate array (FPGA)\renewcommand{\fpga}{FPGA\xspace}\renewcommand{\fpgas}{FPGAs\xspace}\xspace}
\newcommand{\fpgas}{field programmable gate arrays (FPGAs)\renewcommand{\fpga}{FPGA\xspace}\renewcommand{\fpgas}{FPGAs\xspace}\xspace}
\newcommand{\sor}{successive over-relaxation (SOR)\renewcommand{\sor}{SOR\xspace}\xspace}
\newcommand{\fdpcg}{Fourier domain pre-conditioned gradient (FDPCG)\renewcommand{\fdpcg}{FDPCG\xspace}\xspace}
\newcommand{\map}{maximum a-posteriori (MAP)\renewcommand{\map}{MAP\xspace}\xspace}
\newcommand{\lqg}{linear-quadratic-gaussian (LQG)\renewcommand{\lqg}{LQG\xspace}\xspace}

\newcommand{\elt}{Extremely Large Telescope (ELT)\renewcommand{\elt}{ELT\xspace}\renewcommand{\elts}{ELTs\xspace}\renewcommand{\eelt}{European ELT (E-ELT)\renewcommand{\eelt}{E-ELT\xspace}\xspace}\xspace}

\newcommand{\elts}{Extremely Large Telescopes (ELTs)\renewcommand{\elt}{ELT\xspace}\renewcommand{\elts}{ELTs\xspace}\renewcommand{\eelt}{European ELT (E-ELT)\renewcommand{\eelt}{E-ELT\xspace}\xspace}\xspace}

\newcommand{\eelt}{European Extremely Large Telescope (E-ELT)\renewcommand{\eelt}{E-ELT\xspace}\renewcommand{\elt}{ELT\xspace}\renewcommand{\elts}{ELTs\xspace}\xspace}

\newcommand{\dugall}{Durham University generalised adaptive optics laser laboratory (DUGALL)\renewcommand{\dugall}{DUGALL\xspace}\xspace}
\newcommand{\fwhm}{full-width at half-maximum (FWHM)\renewcommand{\fwhm}{FWHM\xspace}\xspace}
\newcommand{\wht}{William Herschel Telescope (WHT)\renewcommand{\wht}{WHT\xspace}\xspace}
\newcommand{\emccd}{electron multiplying CCD (EMCCD)\renewcommand{\emccd}{EMCCD\xspace}\xspace}
\newcommand{\dasp}{Durham \ao simulation platform (DASP)\renewcommand{\dasp}{DASP\xspace}\renewcommand{\thedasp}{DASP\xspace}\xspace}
\newcommand{\thedasp}{the Durham \ao simulation platform (DASP)\renewcommand{\dasp}{DASP\xspace}\renewcommand{\thedasp}{DASP\xspace}\xspace}
\newcommand{\mpi}{Message Passing Interface (MPI)\renewcommand{\mpi}{MPI\xspace}\xspace}
\newcommand{\smp}{symmetric multi-processing (SMP)\renewcommand{\smp}{SMP\xspace}\xspace}
\newcommand{\svd}{singular value decomposition (SVD)\renewcommand{\svd}{SVD\xspace}\xspace}
\newcommand{\gpu}{graphical processing unit (GPU)\renewcommand{\gpu}{GPU\xspace}\renewcommand{\gpus}{GPUs\xspace}\xspace}
\newcommand{\gpus}{graphical processing units (GPUs)\renewcommand{\gpu}{GPU\xspace}\renewcommand{\gpus}{GPUs\xspace}\xspace}
\newcommand{\fft}{fast Fourier transform (FFT)\renewcommand{\fft}{FFT\xspace}\xspace}
\newcommand{\ifu}{integral field unit (IFU)\renewcommand{\ifu}{IFU\xspace}\xspace}
\newcommand{\darc}{the Durham \ao real-time controller (DARC)\renewcommand{\darc}{DARC\xspace}\renewcommand{\Darc}{DARC\xspace}\xspace}
\newcommand{\Darc}{The Durham \ao real-time controller (DARC)\renewcommand{\darc}{DARC\xspace}\renewcommand{\Darc}{DARC\xspace}\xspace}
\newcommand{\cots}{commercial off-the-shelf (COTS)\renewcommand{\cots}{COTS\xspace}\xspace}
\newcommand{\rtcp}{real-time control pipeline (RTCP)\renewcommand{\rtcp}{RTCP\xspace}\xspace}
\newcommand{\rms}{root-mean-square (RMS)\renewcommand{\rms}{RMS\xspace}\xspace}
\newcommand{\sFPDP}{serial Front Panel Data Port (sFPDP)\renewcommand{\sFPDP}{sFPDP\xspace}\xspace}
\newcommand{\wpu}{wavefront processing unit (WPU)\renewcommand{\wpu}{WPU\xspace}\xspace}
\newcommand{\canary}{CANARY\xspace}

\newcommand{\rtcs}{real-time control system (RTCS)\renewcommand{\rtcs}{RTCS\xspace}\renewcommand{\rtcss}{RTCSs\xspace}\xspace}
\newcommand{\rtcss}{real-time control systems (RTCSs)\renewcommand{\rtcs}{RTCS\xspace}\renewcommand{\rtcss}{RTCSs\xspace}\xspace}
\newcommand{\eso}{European Southern Observatory (ESO)\renewcommand{\eso}{ESO\xspace}\renewcommand{\theeso}{ESO\xspace}\xspace}
\newcommand{\theeso}{the \eso\renewcommand{\theeso}{ESO\xspace}\xspace}
\newcommand{\scao}{single conjugate \ao (SCAO)\renewcommand{\scao}{SCAO\xspace}\renewcommand{\Scao}{SCAO\xspace}\xspace}
\newcommand{\Scao}{Single conjugate \ao (SCAO)\renewcommand{\scao}{SCAO\xspace}\renewcommand{\Scao}{SCAO\xspace}\xspace}
\newcommand{\glao}{ground layer \ao (GLAO)\renewcommand{\glao}{GLAO\xspace}\xspace}
\newcommand{\eagle}{ELT Adaptive optics for GaLaxy Evolution (EAGLE)\renewcommand{\eagle}{EAGLE\xspace}\xspace}
\newcommand{\maory}{multi-conjugate \ao relay for the \eelt (MAORY)\renewcommand{\maory}{MAORY\xspace}\xspace}

\title[Real-time simulation for AO]{A real-time simulation facility for
  astronomical adaptive optics}

\author[A. G. Basden]{Alastair Basden$^{1}$\thanks{E-mail:
    a.g.basden@durham.ac.uk (AGB)}\\
$^{1}$Department of Physics, South Road, Durham, DH1 3LE, UK}

\begin{document}
\maketitle



\begin{abstract}
In this paper we introduce the concept of real-time
hardware-in-the-loop simulation for astronomical adaptive optics, and
present the case for the requirement for such a facility.  This
real-time simulation, when linked with an adaptive optics real-time
control system, provides an essential tool for the validation,
verification and integration of the Extremely Large Telescope
real-time control systems prior to commissioning at the telescope.  We
demonstrate that such a facility is crucial for the success of the
future extremely large telescopes.
\end{abstract}


\section{Introduction}
All ground-based astronomical telescopes perform science by observing
through the Earth's atmosphere, which has a degrading effect on the
images obtained in the optical and near infrared.  \Ao
\citep{adaptiveoptics} is a technology employed on most major
telescopes which seeks to remove some of the effects of atmospheric
turbulence, producing clearer, high resolution science images as a
result.  It is a crucial technology for the next generation \elt
facilities \citep{eelt} which will spend the significant majority of
their time producing \ao corrected observations.

\subsection{The ELT AO system integration problem}
The design process of an \ao system involves extensive numerical
simulation and modelling.  At the scale of the systems required for
the \elts, this modelling is an extremely time consuming process using
currently available tools.  It can often take many hours to cover a
single point of the large explorable parameter space, making
responsive design decisions difficult.

Once these systems have been designed and the components fabricated,
verification and integration with \elt facilities is required,
followed by instrument commissioning.  Herein lies a significant
problem: The \elt \ao systems will be dependent on major components
that form part of the \elt design, for example the M4 mirror for the
\eelt, and an extensive \lgs launch capability, as well as a large
number of expensive, technically advanced \wfss.  Not only are these
components expensive, they are often also physically large, and so
using them during laboratory \ao system integration would result in
huge cost and complication.  The complexities of \elt-scale \ao
systems means that the design and build of these systems is likely to
take place at multiple sites around the world.  Therefore, if multiple
copies of components are required (including dummy components, for
example a non-deformable mirror of an equivalent size), this will also
greatly add to the cost of the instruments.  The wide-field, laser
assisted \ao systems proposed for operation with most \elt instruments
also require large numbers of fast, low noise wavefront sensors.
These state-of-the-art components are expensive, and likewise it would
not be possible to duplicate them across every \ao laboratory involved
with the design and build of the \ao system in question.  The \rtcs
which provides \dm commands in response to \wfs inputs requires these
components to be present so that optical and electrical feedback loops
and calibration procedures can be implemented and tested.  In
addition, this \rtcs is integral to the end-user tools required for
\ao system operation, and therefore is required for the development
and testing of these tools.  Subsystem integration at the telescope
itself is also not a solution due to the high costs of \elt time.

In this paper, we develop a solution for this currently unsolved
verification, integration and commissioning problem for \elt \ao
instruments.  We also comment on the additional benefits that this
solution will bring to the \ao community.  In \S2 we introduce the
concept of a real-time hardware-in-the-loop simulation capability
focused on enabling test, verification and integration of \elt \ao
systems.  In \S3 we consider case studies where we have used a
simulation to real-time control link, and where hardware-in-the-loop
simulation is essential.  We conclude in \S4.

\section{Real-time AO simulation}
\begin{figure}
\centering\includegraphics[width=\linewidth]{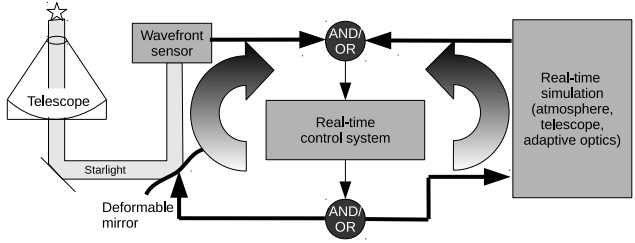}
\caption{A diagram showing the concept of a real-time simulation
  capability.  The real-time control system is able to use a mixture
  real and modelled components.}
\label{fig:rtsim}
\end{figure}

The impracticalities associated with using the large physical
components of an \elt \ao system for verification and integration lead
to the conclusion that simulation of these components is necessary.
On the one hand, these components could be replaced with a physical dummy
version that simply provides or accepts the equivalent electronic
signals.  This however only allows interface testing, and does not
allow full system testing, for example, of:
\begin{enumerate}
\item Algorithms within the \rtcs.
\item Calibration procedures for performance optimisation.
\item Closed loop latency, bandwidth and jitter tests.
\item Real-time response of the \rtcs.
\item Stress-testing of the system under typical usage patterns.
\item Interaction with telescope facilities.
\end{enumerate}
Therefore, a hardware-in-the-loop simulation model of these components
is also required (Fig.~\ref{fig:rtsim}).  Not only must these
simulation models interface to the \rtcs, they must also interact with
any physical components present, and must be able to operate at
real-time rates to allow proper testing of the \rtcs subsystems.  This
facility will also be useful for \ao systems on existing telescopes,
and is therefore not restricted to \elt systems.

\Ao simulation, using Monte-Carlo techniques, is routinely used to
model the performance of \elt \ao systems
\citep{basden12,2012SPIE.8447E..23W}.  These models, which include the
effect of atmospheric turbulence, sensor noise and physical models for
\wfs cameras, \dms and science cameras are highly computationally
intensive.  Efforts have been made to use non-conventional computing
hardware \citep{basden4,gratadour} to increase the speed of \ao
simulations, thus reducing the time taken to explore a given parameter
space.  Additionally, recent advances in conventional computing
technologies, including \gpus and many-core technologies increase the
potential for a further reduction in simulation execution times.
However, to allow an \elt-scale \ao simulation to reach real-time
rates would require a substantial computational hardware investment,
which may not always be possible or appropriate.  We therefore propose
a five step plan for development of a real-time hardware-in-the-loop
\ao simulation capability, outlined in the following sections.  Each
of these steps should be considered as a solution to a subset of
problems in its own right, and it is not necessary to implement all
steps for every instrument or at every laboratory, allowing a cost and
complexity trade-off to be made.

It is important to note that this real-time hardware-in-the-loop
simulation facility is not designed with the goal of high fidelity \ao
simulation, or for parameter space investigation or new algorithm
development.  Rather it is aimed at solving the \ao system integration
issues when faced with complicated telescope interfaces and components
not present during laboratory integration, to reduce the risks
associated with \ao system development.

\subsection{Interfacing of simulation with a real-time control system}
The first step to be taken towards solving the \elt \ao verification,
integration and commissioning problem is to interface a full
Monte-Carlo simulation code with the \ao \rtcs.  This should be
implemented in a way which allows the \rtcs to be blind to the fact
that it is operating in simulation mode, i.e.\ all the \rtcs
algorithms should be as intended for on-sky use.  This is essential
because it allows algorithms that are not implemented within typically
Monte-Carlo simulations, yet which greatly improve on-sky performance,
to be fully investigated and tested, such as adaptive windowing
techniques and brightest pixel selection \citep{basden10}.

\begin{figure}
\includegraphics[width=\linewidth]{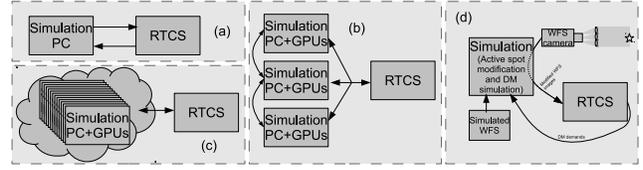}
\caption{A diagram demonstrating some of the stages of
  hardware-in-the-loop real-time simulation.  (a) Step one, a
  simulation interfaced with a RTCS.  (b) Step two, a fast simulation
  interfaced with a RTCS.  (c) Step three, a real-time simulation
  interfaced with a RTCS.  (d) Step four, physical component
  interchange and modelling (here showing a missing WFS, spot
  modification of a physically present WFS, and a missing DM).}
\label{fig:simStages}
\end{figure}

At this step, the simulated components will not be operated at
real-time rates, allowing conventional PC hardware to be used for the
simulation, without requiring massive parallelisation techniques.
This will allow the \ao loop to be engaged within the \rtcs, and
performance metrics obtained.  Costs will be minimised, and so this
step is appropriate for wide distribution to laboratories involved
with \ao system development for who real-time operation is not
essential.  This is demonstrated in Fig.~\ref{fig:simStages}(a).  We
suggest that this step is particularly appropriate for \ao systems on
10~m class telescopes where reasonable simulation frame rates are
achievable on modest computational hardware.

We have demonstrated this step by interfacing \thedasp \citep{basden5}
with \darc \citep{basden9,basden11}, which has provided an ideal tool
for developing and testing on-sky algorithms used in the CANARY
instrument \citep{canaryresultsshort}.  In this case, the link between
simulation and \rtcs was implemented using Ethernet sockets rather
than the \sFPDP communication used by the \wfs cameras.  Thus, to
change between real and simulation mode, it is necessary to load and
unload the relevant \darc modules within the \rtcs.  However, due to
the design of \darc, such modularity is trivial, and no algorithm
changes are required.

In this first step of real-time simulation, it should be noted that there
are no physical components.  The atmosphere, telescope, \wfss, optics
and \dms are all simulated.  For a \rtcs that can operate entirely
within a PC (such as \darc, when using appropriate modules), this step
allows the whole \rtcs and simulation to be operated on a single PC,
providing ultimate flexibility, suitable for duplication by many
developers simultaneously.

In addition to providing a test harness for the \rtcs with all the
hardware components (albeit simulated), this real-time simulation
interface has a further advantage over laboratory test-bench
demonstration: The ability to explore a far wider atmospheric
turbulence parameter space.  This includes the simulation of a far
greater number of atmospheric phase screens than are available with a
typical test bench.  We have simulated up to 40 phase screens at
\elt-scale using \dasp, while a typical laboratory setup will contain
up to four screens \citep{dragon}.  This provides the ability to model
velocity dispersion within layers, and also to model layers with
finite thickness, allowing real-time implementations of key algorithms
to be tested.

\subsubsection{Playback of images and slopes}
At this step, we include the ability to replay pre-generated \wfs
images into the \rtcs, and also the possibility of replaying pre-recorded
wavefront slope measurements, should the \rtcs be able to accept
this.  This would allow some validation (particularly of some wavefront
reconstruction algorithms) to be implemented by comparing \rtcs output
with the expected output.

\subsection{Fast simulation}
The next step is for moderate acceleration of the simulation code,
using hardware that is readily available and affordable, comprised of
typically a small number of \gpu accelerated computer servers as shown
in Fig.~\ref{fig:simStages}(b).  Although this adds to the complexity
of the simulated system, it has the advantage that the system update
rate will be up to a few tens of Hertz for an \elt-scale instrument,
allowing users to control the \rtcs without unacceptable delays, and
to view telemetry data (\wfs images, reconstructed phase, etc.) at a
rate that is acceptable to the human eye, allowing users to better
appreciate how the \ao system will operate as a whole, and providing a
reasonable responsiveness to user interface controls.

\subsection{Real-time simulation}
To achieve real-time rates for an \ao simulation and \rtcs combination
at \elt-scale, greater computational resources will be required,
consisting of a moderate computing cluster as shown in
Fig.~\ref{fig:simStages}(c).  Here we provide an estimate of the
required computing power for a typical \elt instrument reminiscent of
the proposed \eagle instrument \citep{eagleScience}, and consider the
hardware that would be required to implement a real-time simulation
capability.  We do not consider here the requirements for the \rtcs
itself, as this has been covered elsewhere \citep{basden11}.

A real-time simulation harness for an \elt \rtcs is essential to solve
the verification, integration and test problem that has been
identified for the \elts.  This will provide a facility to allow the
full \rtcs to be integrated with \elt systems prior to arrival at the
telescope.  

\subsubsection{Simulation components}
A real-time simulation facility must model many separate components so
that a realistic test harness for the \rtcs can be provided.  These
include as a bare minimum:
\begin{enumerate}
\item Wavefront phase distortions caused by layers of atmospheric
  turbulence (up to 40 layers are required for accurate modelling, \citet{2012SPIE.8447E..57C})
\item The integrated phase distortions along given lines-of-sight
\item Telescope and \wfs optics (including \lgs spot elongation)
\item \wfs noise 
\item \dms and associated optical components
\item Science cameras for performance verification
\end{enumerate}

Since the real-time simulation must operate for undetermined periods
of time, we assume that a technique for generating infinitely long
atmospheric phase screens will be used \citep{assemat}, based on the
statistical co-variance of the turbulent phase.

\subsubsection{Computational complexity}
We now consider the minimum computational requirements that will be
required for an \elt-scale real-time simulation of an \ao instrument
operating at 250~Hz (that of \eagle).  An estimate for the required
operations are given in table~\ref{tab:operations} for a single
line-of-sight, turbulent layer or \wfs.  However, it should be noted
that this will vary depending on simulation input parameters, such as
number of turbulent atmospheric layers, layer heights, wind
velocities, \wfs pixel scale and many other factors, so should only be
treated as representative.  Additionally, we have only considered the
basic algorithms required, and it is likely that a true real-time
simulation would require extra algorithms to improve fidelity.  We
also assume that data accesses are for data that is contiguous in
memory, or that stepped memory access is available (as with the Intel
Sandy Bridge processors).  The required operations match those that we
currently use in \dasp.

\ignore{Assume O(1) for each random number generation.}

\begin{table*}
\begin{minipage}{17cm}
\caption{A table detailing the operations required for
  real-time simulation.  Where available, operations are given as
  standard BLAS function names.  Memory access is given in 4-byte
  units.  $N$ is the number of phase pixels across the telescope pupil
  (~1600 for \eagle), and $M$ is the number of sub-apertures across
  the telescope pupil (~80 for \eagle).}
\label{tab:operations}
\begin{tabularx}{\linewidth}{p{1.9cm}|X|p{1.2cm}|p{1cm}|p{1.5cm}}
\hline Algorithm & Operations & Complexity & Memory access & EAGLE at 250~Hz \\ \hline 

Phase screen generation & Per layer:
2 GEMV, 1 AXPY & $16N^2+2N$ & $8N^2+2N$ & 10~GFLOPS,
20~GBs$^{-1}$\\ 

Line-of-sight integration & Per layer and direction:
2D-spline interpolation, AXPY & $N^2\ln N + N^2$ & $2N^2$ &
5~GFLOPS, 4~GBs$^{-1}$\\ 

\wfs model & Per \wfs: 2 AXPY, Cos/Sin, 2D FFT, 2D convolution, AXPY,
Noise addition & $13N^2 + 3N^2\ln M$ & $5N^2$ & 16~GFLOPS,
12~GBs$^{-1}$ \\

\dm model & Per \dm and direction: 2D-spline interpolation, AXPY &
$N^2\ln N + N^2$ & $2N^2$ & 5~GFLOPS, 4~GBs$^{-1}$\\

Science & Per science direction: 3 AXPY, Cos/Sin, 2D FFT, SUM &
$10N^2+N^2\ln N$ & $6N^2$ & 11~GFLOPS, 15~GBs$^{-1}$ \\

\hline
\end{tabularx}
\end{minipage}
\end{table*}

Using these computational complexity estimates, we can place an order
of magnitude estimate on the simulation computational requirements for
\elt \ao systems, as shown in table~\ref{tab:estimates}.  If we assume
that the real-time simulation is to be implemented using \gpu
technology, then estimates can be placed on the size of a system
required to implement this real-time simulation.  The current
generation of \gpus, such as the NVIDIA GTX-780 can reach
approximately 4~TFLOPS of single precision floating point performance,
and have a theoretical internal memory bandwidth of up to
250~GBs$^{-1}$.  This internal memory bandwidth is therefore the
limiting factor in real-time simulation performance.  If we assume
that for mixed algorithms, 50\% of the theoretical bandwidth peak can
be reached \citep{basden11}, then three of the cases in
table~\ref{tab:estimates} are achievable using 18 \gpus.  The most
demanding case (a 40 layer simulation) would require about 50 \gpus.  A
suitable system would contain several PCs to host these \gpus, and
thus require inter-node communications.  This introduces additional
complexity to the system, requiring time for the transmission and
marshalling of data.  We therefore suggest that additional computing
power is required to reduce computation time, thus allowing additional
time for data communications (which we do not cover here).  Some
overhead is also necessary to allow for communication with the \rtcs.
A factor of two would seem reasonable, requiring a 36 \gpu system to
obtain real-time simulation rates for models with ten atmospheric
layers.

The \elt simulation problem is highly suitable for parallelisation,
and will benefit from improvements made in future computational
hardware.  Not only is it possible to parallelise at the component
level (\wfss, phase screens, \dms etc.), but it is also possible to
split computation of many components across different computational
hardware units, with clean partitioning between units requiring little
or no inter-unit communication.  As an example, \wfs sensor simulation
can be divided easily on a per-sub-aperture basis, producing sections
of \wfs images on separate \gpus before marshalling to produce the
final image to be sent to a real-time control system.  Such
marshalling is trivial when \gpus are on the same host, and requires
additional network bandwidth when generated on separate hosts.  
Currently, up to eight \gpus can be used with a single host on
commonly available motherboards.  

Almost all parts of \ao simulation can either be parallelised in this
way or can be generated identically in different parts of the
simulation hardware where there is potential to increase simulation
speed by reducing network bandwidth requirements.  For example it may
be preferable to generate multiple instances of the same atmospheric
phase screen, rather than distributing one instance to all the
simulation components that require it.

\begin{figure}
\centering\includegraphics[width=\linewidth]{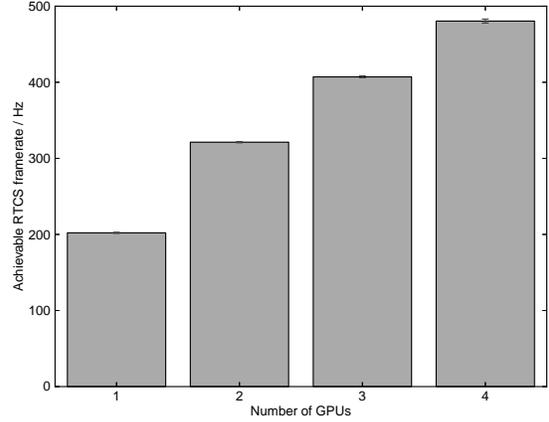}
\caption{A figure showing scaling of real-time control system
  performance as a function of number of GPUs used, based on results
  obtained using the \darc system configured for an \elt case.}
\label{fig:gpuscaling}
\end{figure}

Scaling of algorithms across multiple \gpus is a well known technique,
which in the case of some well suited algorithms, provides performance
improvements almost proportional to the number of \gpus used.  In
typical real-world applications, scaling is slightly worse.
Figure~\ref{fig:gpuscaling} shows performance scaling with number of
\gpus of the \darc \rtcs configured for a $84\times84$ sub-aperture
\ao system with 3 \dms (a total of 9700 actuators).  Here, the \gpus
were used for wavefront reconstruction, and although algorithms used
for simulation are different, we take this as representative of
real-world algorithms used for \ao.  In this case, \wfs calibration
and slope calculation were performed in \cpu which will also have some
effect on the system scaling.  The scaling that we achieve is similar
to that reported for \gpu accelerated \ao simulation
\citep{gratadour}, with performance scaling slightly worse than
proportional to the number of \gpus present.  

\begin{figure}
\includegraphics[width=\linewidth]{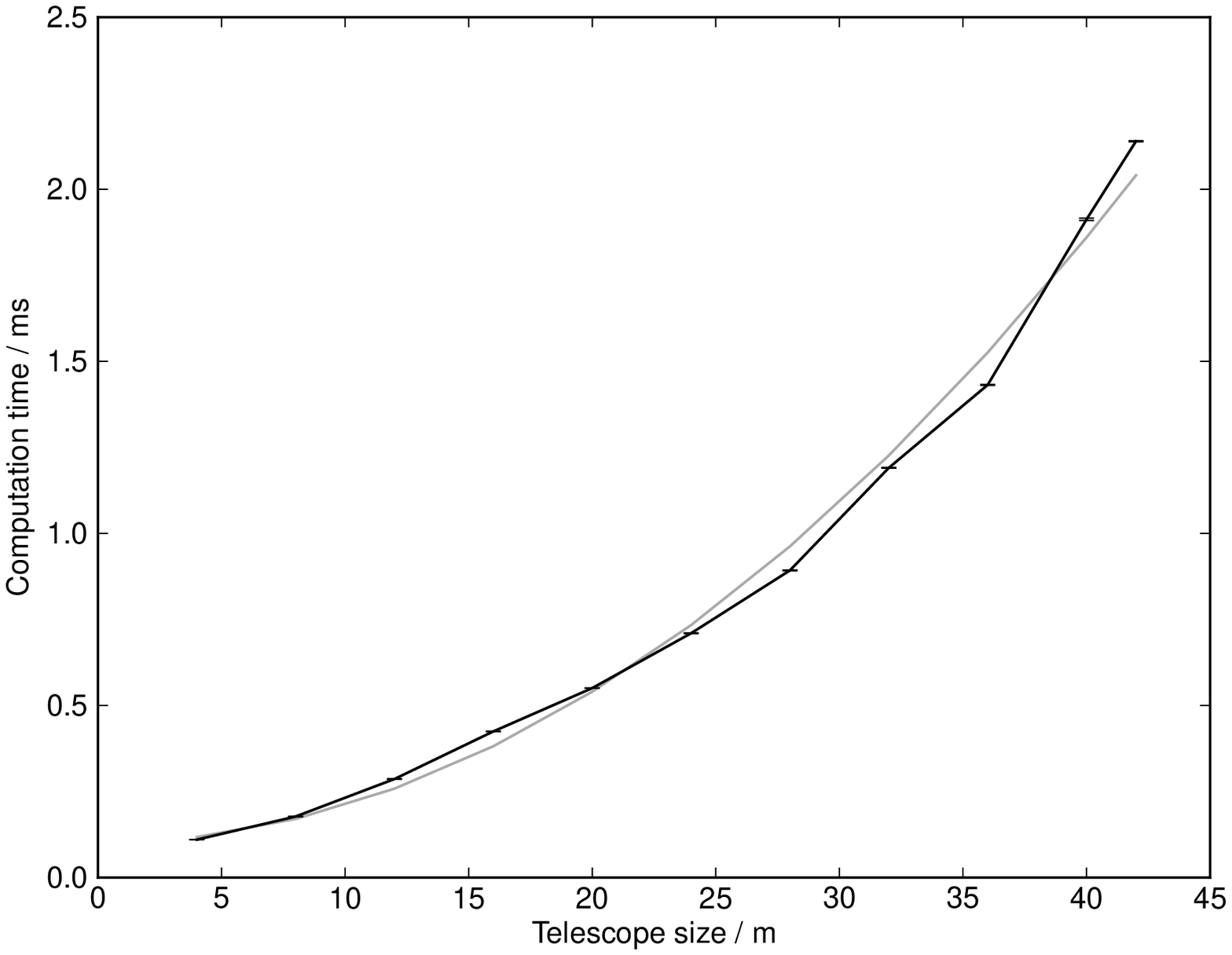}
\includegraphics[width=\linewidth]{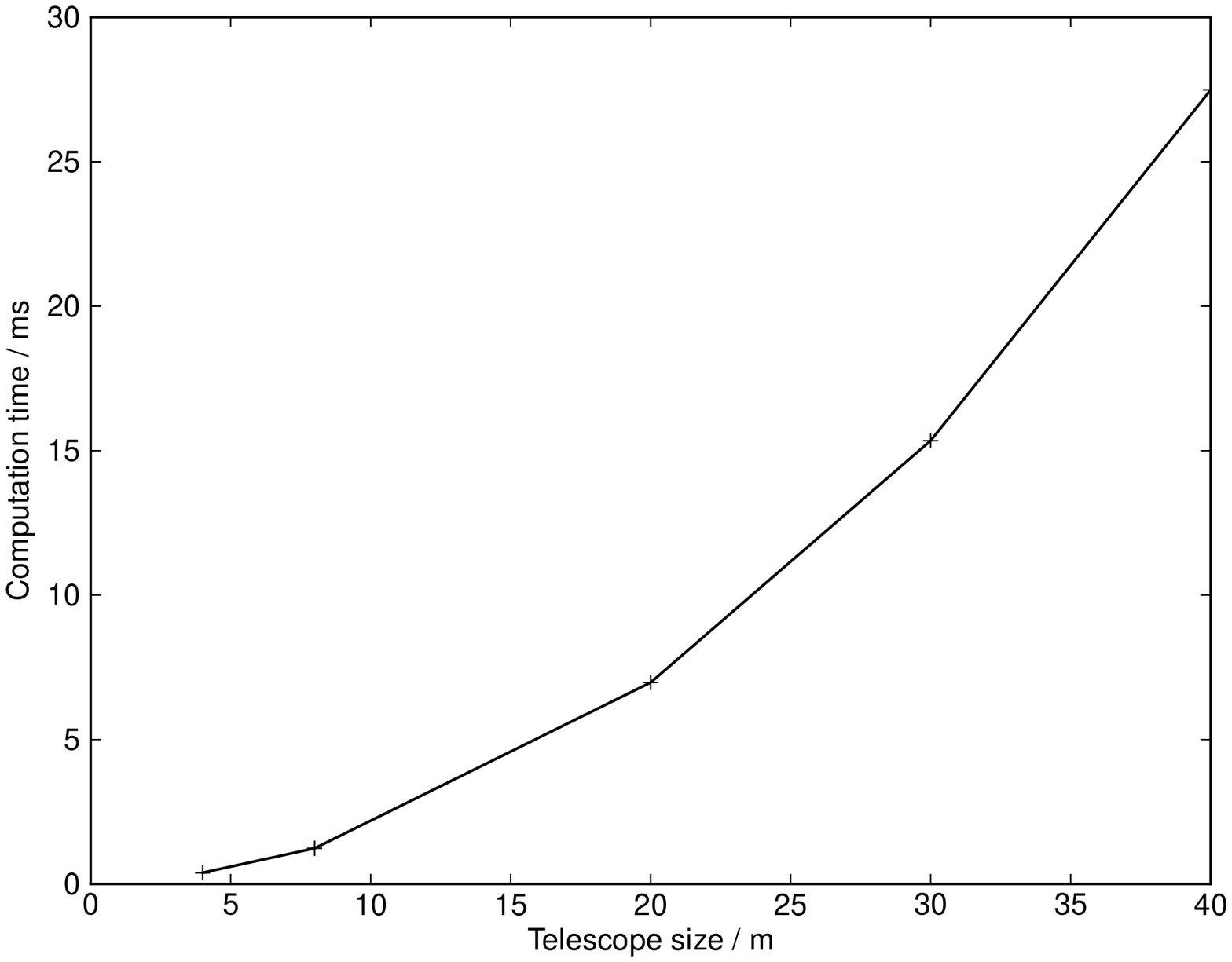}
\caption{(a) A figure showing scaling of real-time control system
  performance as a function of telescope size using a single GPU for
  wavefront reconstruction, assuming 0.5~m sub-apertures.  A quadratic
  fit is shown in grey.  (b) Showing
  scaling of AO simulation as a function of telescope size, using data
  from \citep{gratadour}.}
\label{fig:systemscaling}
\end{figure}

The theoretical scaling of computational requirements as a function of
\ao system size can be obtained from table~\ref{tab:operations}.  We
have investigated \rtcs performance scaling using \darc with
reconstruction performed using a single \gpu, and other operations
performed on the host processor.  Results are shown in
Fig.~\ref{fig:systemscaling}(a), and although these algorithms are
different from those that would be required in simulation, demonstrate scaling
with system size.  Previous studies of \scao simulation
\citep{gratadour} have investigated scaling of simulation with \ao
system order on \gpu, and from this information we have computed
simulation time (excluding slope estimation and wavefront
reconstruction), as shown in Fig.~\ref{fig:systemscaling}(b).  This
includes most of the algorithms required for hardware-in-the-loop
simulation, demonstrating a scaling proportional to square of
telescope diameter, in agreement with table~\ref{tab:operations}.

Random access memory (RAM) in the current generation of \gpus is
limited to about 6~GB, while the \elt scale simulations that we use
require up to 32~GB RAM.  A real-time simulation facility would be
spread across many \gpus, and so we do not foresee any memory problems
arising, since the simulation can be naturally partitioned and
parallelised in such a way that memory consumption is spread out
between the \gpus.  In the fast (non-real-time) simulation case, fewer
\gpus will be used, possibly containing less than 32~GB between them.
However, the idea in this case is that the \gpus are then used for
offloading parts of the simulation computation, not all of it, and so
available memory is likely to be sufficient.  Future generations of
\gpus and many-core processors are likely to have more memory (for
example the Intel Xeon Phi has 16~GB), further alleviating any memory
problem.

\begin{table*}
\begin{minipage}{17cm}
\caption{A table estimating real-time simulation
  computational requirements for different proposed \eelt instruments.}

\label{tab:estimates}
\begin{tabularx}{\linewidth}{XXXp{1cm}p{1.5cm}p{3.5cm}p{3cm}}
\hline Instrument & \# layers & \# \wfs & \# \dms & Frame rate
& Requirements & Ref \\ \hline

\ignore{10*10+5*31*10+16*11+5*20+11*20}
\ignore{20*10+4*31*10+12*11+4*20+15*20}
EAGLE & 10 & 11 & 20 & 250 & 2.1~TFLOPS, 2.0~TBs$^{-1}$ & \citep{basden12}\\

\ignore{10*40+5*31*40+16*11+5*20+11*20} 
\ignore{20*40+4*31*40+12*11+4*20+15*20}
EAGLE & 40 & 11 & 20 & 250 & 7.1~TFLOPS, 6.3~TBs$^{-1}$ & \citep{basden12}\\

\ignore{(10*10+5*12*10+16*9+5*3+9*1)*500/250  - 1 science direction}
\ignore{(20*10+4*12*10+12*9+4*3+15*1)*500/250}
MAORY & 10 & 9 & 3 & 500 & 1.7~TFLOPS, 1.6~TBs$^{-1}$ & \citep{2010SPIE.7736E..99F}\\

\ignore{(10*40+5*12*40+16*9+5*3+9*1)*500/250  - 1 science direction}
\ignore{(20*40+4*12*40+12*9+4*3+15*1)*500/250}
MAORY & 40 & 9 & 3 & 500 & 5.9~TFLOPS, 5.7~TBs$^{-1}$ & \citep{2012SPIE.8447E..57C}\\

\ignore{(10*10+5*9*10+16*8+5*1+11*1)*800/250}
\ignore{(20*10+4*9*10+12*8+4*1+15*1)*800/250}
ATLAS & 10 & 8 & 1 & 800 & 2.2~TFLOPS, 2.2~TBs$^{-1}$ & \citep{2010aoel.confE2002F}\\

\ignore{(10*40+5*9*40+16*8+5*1+11*1)*800/250}
\ignore{(20*40+4*9*40+12*8+4*1+15*1)*800/250}
ATLAS & 40 & 8 & 1 & 800 & 7.5~TFLOPS, 7.5~TBs$^{-1}$ & \citep{2012SPIE.8447E..57C}\\

\hline
\end{tabularx}
\end{minipage}
\end{table*}

Although our treatment of computational requirements has been
preliminary, we have nevertheless been able to show that an \elt-scale
real-time simulation capability is achievable using existing
computational hardware, though our estimates for hardware required
are order-of-magnitude only.

\subsubsection{Accuracy and performance trade-offs}
In order to achieve real-time rates on available hardware, it may be
necessary to reduce accuracy of the simulations.  This will have some
impact on the \ao system performance, depending on what
simplifications are made.  However, since this real-time simulation
facility is focused on \elt system integration rather than high
fidelity simulation this is unlikely to cause problems in most
situations.  When high fidelity simulations are required, the
simplifications can be removed, resulting in a fast (but not
real-time) \ao simulation capability.

An example of an accuracy and performance trade-off that can made is
that of telescope pupil sampling for phase-screen generation.
Reducing the sampling reduces both the computational requirements for
the simulation, and also the accuracy of the simulation.  Another
example is the number of atmospheric layers modelled.  Fewer layers
would result in reduced simulation fidelity, though with lower
computational cost.  How far each trade-off can be taken will depend
on the particular circumstances under investigation.

\Dm modelling fidelity is also a trade-off that can be made to reduce
computational complexity.  Simplified models for \dms will result in
less accurate simulations though enable real-time rates with reduced
hardware requirements.  Similarly, wavefront sensor models can also be
simplified at the expense of accuracy, for example by ignoring
vignetting, and by using pre-generated, or simplified random noise
generators to model detector readout noise.

Elongation of \lgs spots is also another area where simulations can be
simplified to reduce computational requirements, including sodium
profile sampling and the size of resulting sub-aperture images.  

Single precision floating point operation is sufficient for almost all
aspects of \elt simulation, and is what we currently use for our \ao
simulations.  The exception is for infinite phase-screen generation
which requires double precision accuracy to maintain valid statistics.

\subsection{Physical component interchange and modelling}
So far we have considered only cases where there are no physical
components present, i.e.\ all such items, including \dms and \wfss are
modelled in simulation.  However, an additional step can also be taken
to allow physically present components to be used with the \rtcs, and
absent components to be modelled.  Examples include systems with a
sub-set of \wfss present, or systems with one or more \dms absent, as
illustrated in Fig.~\ref{fig:simStages}(d).

The combination of physical and modelled components introduces an
extra degree of complexity for the modelling, and thus increased
computational requirements if real-time rates are to be maintained.

\subsubsection{Absent wavefront sensors}
Let us first consider the case where not all \wfss are present at the
\ao system integration laboratory.  This situation is particularly
likely to arise for wide-field tomographic \ao systems which require a
large number of high-speed, low noise \wfss, with cost preventing
replication of the complete set of \wfss at all integration
facilities.  It is therefore necessary to use hardware-in-the-loop
simulation of the missing \wfss, to enable \rtcs integration.  If the
atmospheric turbulence is deterministic, for example it is created
using a system of rotating phase screens or a set of liquid crystal
screens, then accurate simulation models can be created.  It is
possible to determine exactly what wavefront aberrations are being
introduced (if necessary by correlating the expected wavefront with
reconstructed phase from the physically present \wfss), thus allowing
simulation of the corresponding non-present \wfss, which would then
deliver \wfs images to the \rtcs almost identical to images that would
have been produced if the physical \wfs had been present.  The \rtcs
can then be used to perform standard tomographic reconstruction and
\dm control, as if all physical \wfss were present.

In the case where the atmospheric turbulence is not deterministic,
this technique cannot be used.  Depending on the requirements for the
system (whether tomographic reconstruction and science verification is
required), other techniques may be possible, for example duplicating
\wfs information (using copies of images from physically present \wfss
to model the absent \wfss), or by using the physically present \wfss
to perform a tomographic wavefront reconstruction, from which the
absent \wfss can be simulated.

For close-loop \ao systems, the real-time simulation code can also be
informed about the shape of the \dms, thus allowing the simulated
\wfss to also respond to \dm surface changes.

\subsubsection{Absent deformable mirrors}
We now consider the case where all \wfss are present, but one or more
\dm is not, for example the \eelt M4 mirror, which is physically large
and not well suited to laboratory integration.   We assume that the
missing \dms are before the \wfss in the optical path, i.e.\ changes
to the \dm surface are measurable using the \wfss.  If this is not the
case (e.g.\ some \dms in an open-loop \moao system) the problem is
actually easier to solve, affecting science verification only.

\begin{figure}
\centering\includegraphics[width=\linewidth]{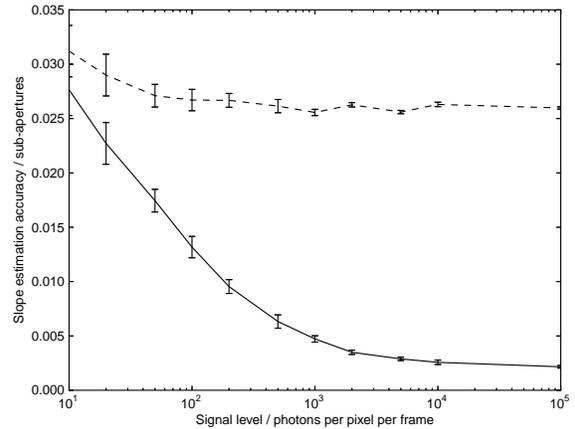}
\caption{A figure showing the accuracy of the active spot modification
  technique.  The curves show the RMS difference between slopes measured
  when a DM is not present but simulated, to slopes measured when a DM
  is present.  The solid curve represents Shack-Hartmann images
  modified using the active spot modification, while the dashed curve
  represents Shack-Hartmann images through turbulence alone
  (i.e.\ without \dm simulation, and hence no spot modification).}
\ignore{A figure showing the accuracy of the active spot modification
  technique.  Wavefront sensor slope measurements are given for each
  active sub-aperture of a $10\times10$ sub-aperture wavefront sensor.
  The black curve shows slope measurements when the input phase to the
  wavefront sensor contains atmospheric and DM contributions.  The
  grey curve shows slope measurements when wavefront sensor input
  phase is atmospheric only, and the sub-aperture spots are convolved
  with those obtained using the \dm shape information.  The agreement
  is good.}
\label{fig:slopes}
\end{figure}

An accurate model of the desired \dm surface shape can be obtained
using the \rtcs outputs, and the amount of detail in these models (for
example assuming a perfect \dm or including hysteresis,
non-linearities and mis-alignments) will depend on circumstances and
requirements.  It is then possible to model how this optical surface
would affect the physical Shack-Hartmann \wfs images relative to an input plane wave,
allowing individual sub-aperture \psfs to be computed.  A convolution
of these \psfs with the real \wfs images then yields a good
approximation to the \wfs images that would have been obtained when a
physical \dm was present as shown in Fig.~\ref{fig:slopes}.  The
presence of \wfs noise reduces the fidelity of this approximation,
though in an integration laboratory, \wfs signal levels can usually be
increased, with statistical noise added back into the real-time
simulation images after this correlation.  These modified \wfs images
are then used as input for the \rtcs, and we call this process
``active spot modification''.

The key benefit of this technique is that it allows the whole \ao
\rtcs to be tested in the absence of a small number of critical components.

\subsubsection{Active spot modification}
The ``active spot modification'' technique allows \rtcs testing of
extended \wfs \psfs in laboratory situations where such \psfs are
difficult to generate optically, for example for elongated laser spots
\citep{2008SPIE.7015E.129L}.  The key concept is to take a closed-loop
laboratory \ao system and modify wavefront sensor images on a
per-sub-aperture basis to allow testing of algorithms within the
\rtcs.  This active modification can include the addition of simulated
photon shot noise, variations in signal intensity, and detector
readout noise.  Investigation of \ao performance scaling with signal
level can be carried out (allowing finer changes in signal than can be
achieved using neutral density filters for example), and also allows
the effect of rapid signal level changes on \rtcs performance to be
investigated.

To use this technique with maximum effect, the true (laboratory)
illumination level needs to be sufficiently bright to be in the high light level
regime (essentially noiseless), which in a laboratory situation, is
usually possible using bright sources.  Similarly, the \wfs detector
pixels should be Nyquist sampled for this technique to work well.  

Figure~\ref{fig:slopes} was generated using a simulation of the
atmosphere and a \dm on a $10\times10$ sub-aperture \ao system on a
4.2~m telescope.  The modelled \wfs was assumed to have 0.1 electrons \rms
readout noise, and a standard centre of gravity algorithm was used to
compute wavefront slope.  

A random shape was applied to the surface of the $11\times11$ actuator
\dm, the surface shape of which was obtained using cubic spline
interpolation.  An atmospheric phase screen was generated using a Von
Karman spectrum \citep{vonkarman} with an outer scale of 30~m, and
Fried's parameter \citep{fried} of 20~cm.

The case for an \ao system with all physical components present was
modelled by producing Shack-Hartmann \wfs images with wavefront phase
modified by both the atmosphere and the \dm.  The wavefront slopes
were then obtained by applying a standard centre of gravity algorithm
to these spot images.  This is equivalent to starlight passing first
through the atmosphere, and then reflected by the \dm before being imaged on
the \wfs.

We then investigated the ``active spot modification'' technique by
producing Shack-Hartmann \wfs images with wavefront phase introduced
by the atmosphere only (and including wavefront sensor noise).  These
images were then modified using the ``active spot modification''
technique, and a centre of gravity algorithm applied to the resulting
images to give the slopes corresponding to a hardware-in-the-loop
simulated \dm.

The difference between slope measurements with the \dm present, and
slope measurements with the hardware-in-the-loop simulated \dm were then
computed, and the \rms difference of all sub-apertures, over many
frames, was used as the metric for performance comparison here.

We also considered the case where the \dm was not present and no
effort made to simulate it, i.e.\ computed slope measurements
represent those of the atmosphere only.  These were again compared
with the slope measurements measured with the \dm present, so that the
benefit of the ``active spot modification'' technique can be seen.

In Fig.~\ref{fig:slopes}, it is clear that the ``active spot
modification'' technique performs well at light levels as low as about
1000 photons per sub-aperture per frame, with reduced performance for
fainter levels.  Such signal levels are easily achievable in
integration laboratories where this technique will be applied.
``Active spot modification'' also always represents an improvement in
slope estimation accuracy when compared with the unmodified \dm-absent
case.

\subsubsection{Absent elongated laser guide stars}
The creation of elongated \lgs spots using laboratory is difficult,
though not impossible \citep{dragon}.  In the absence of elongated
spots, we can simulate this using the active spot modification
technique outlined in the previous section.  A typical laboratory
arrangement in this case would be to use physical \wfss and
\dms, but with \wfss imaging point sources.  The real-time simulation
would then be used to modify the \wfs images by convolving each
sub-aperture image with an appropriate elongated \psf, before passing
to the real-time control system.  Additionally, the simulation could
also be used to introduce photon shot noise and detector readout
noise.

\begin{figure}
\includegraphics[width=0.3\linewidth]{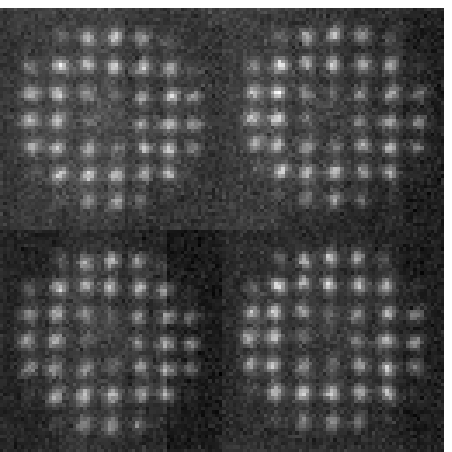}%
\includegraphics[width=0.3\linewidth]{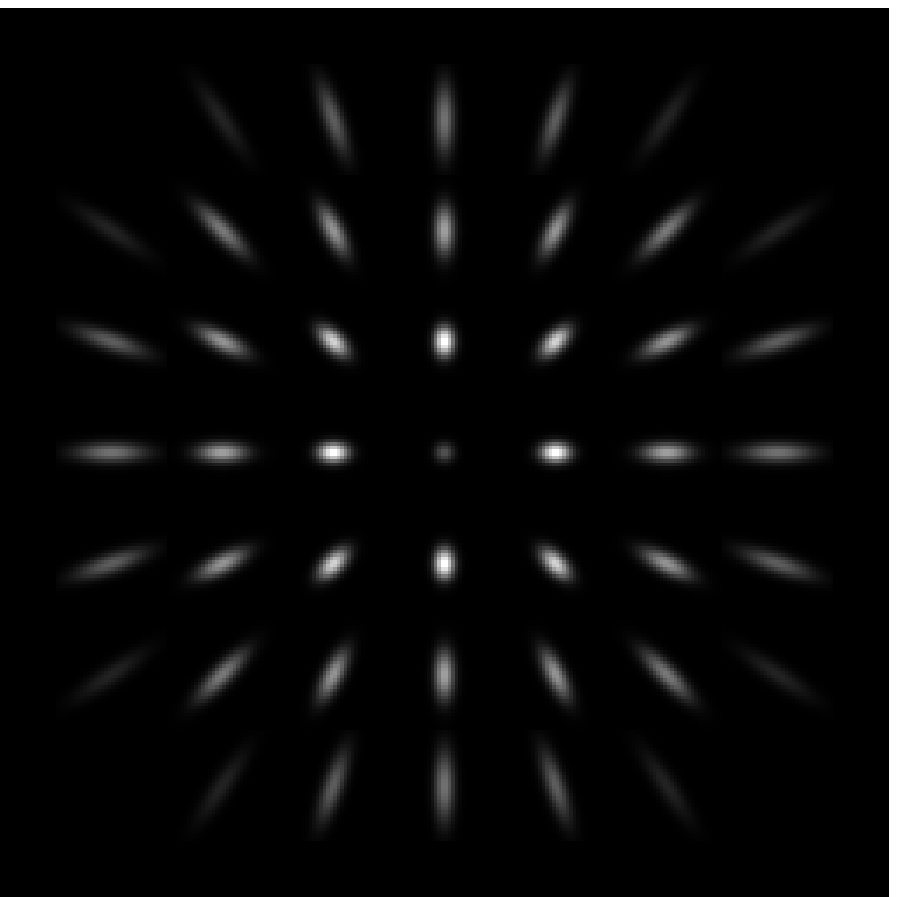}%
\includegraphics[width=0.3\linewidth]{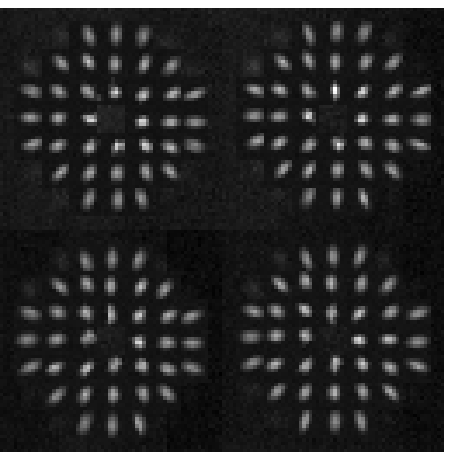}
\caption{A figure demonstrating active spot modification within the
  DARC RTCS.  (a) Raw images of (unelongated) spots generated with
  light sources on a bench, within the CANARY sky simulator.  (b) The
  spot LGS elongation pattern used with active spot modification.  (c)
  Actively modified SHS spots, which are treated by the real-time
  control system as the raw images, including spot elongation, photon
  shot noise and readout noise.  The elongation is clear, allowing
  RTCS algorithms such as correlation WFSing to be tested.}
\label{fig:spotsModified}
\end{figure}

This active spot modification is a feature available in \darc, where
we currently use it to modify \wfs spot \psfs and \wfs noise levels on
\canary, as shown in Fig.~\ref{fig:spotsModified}.  We have successfully demonstrated this technique in
closed-loop, therefore verifying this process for use in real-time
simulation.

\subsection{Hard-real-time testing}
A key parameter for a real-time control system is the reliability with
which it is able to compute and deliver \dm commands in response to
\wfs input within a given time period.  All \rtcss built using
non-deterministic hardware (such as \cpus) will suffer from some
degree of uncertainty about the latency between input and output.
This variation in latency is termed the jitter of an \ao system.  

The real-time simulation capability described here will also be
non-deterministic, and so include jitter.  Determining the true jitter
of a \rtcs will be difficult when using a real-time simulation.
Therefore, we also propose that a hard-real-time capability should
also be implemented alongside the hardware-in-the-loop simulation,
with very limited functionality, but with essentially zero jitter,
obtained by using deterministic hardware.

Such a facility, typically developed using \fpga hardware, would
produce a predetermined set of \wfs images with programmable
inter-pixel and inter-row timings matching that of the true \wfs
camera readout sequence.  It should be noted that this facility does
no calculations to produce \wfs images, rather, it simply sends a set
of pre-generated images.  The hardware interface to this facility
would be identical to the \wfss, so that identical \rtcs hardware
could be used.  The \rtcs output would also be captured by this
device, using an identical interface to the \dms.  A cycle-by-cycle
history of latency can then be build up over millions of cycles,
allowing an accurate jitter profile to be produced, by recording the
time at which a \wfs frame was sent to the \rtcs, and recording the
corresponding delay before receiving \dm demands from the \rtcs.

We are currently developing such a system with a 10G Ethernet
interface, allowing it to be used with proposed \eelt instruments, as
well as for in-house testing of \rtcss.

\subsection{Steps to system integration}
Integration of an \elt \ao system can be performed using the stages
outlined above.  Initial fast simulation tools can be interfaced to
the \rtcs allowing \rtcs algorithm testing.  Testing of the real-time
implementations of wavefront reconstruction algorithms can be
simplified at this stage by replaying slopes rather than simulated
images into the \rtcs, and observing the \dm command output.
Additionally, the hard-real-time deterministic image generator can be
used at this stage to demonstrate \rtcs suitability for the task
at hand, by making extensive jitter measurements.  This can be
repeated when new algorithms are added to the \rtcs pipeline.

As physical components become available at the integration laboratory,
they can be added into the simulation loop, replacing simulated
components and allowing a gradual buildup of the \ao system to take
place, until it can be integrated with the telescope, when all
components will be present.

It should also be noted that if the \rtcs is configurable, it will not
always be necessary to test algorithms at full \elt system scale.
Rather, \rtcs algorithm testing can often be carried out on scaled down systems.
However, there will always be some size-specific algorithms which must
be tested at full scale.  Additionally, testing of offloads to telescope
facilities (including guiding and active optics systems) will require
simulation at full system scale.

\section{Real-time simulation case studies}
Having introduced the concept of real-time simulation for \ao, it is
useful to provide some case studies where this facility is or will be
useful, or indeed, essential.  To date, we have used a simulation to
\rtcs bridge corresponding to step one above, and do not yet have a
full real-time hardware-in-the-loop simulation facility as described
in this paper.

\subsection{Advanced wavefront reconstruction investigations}
The \canary \ao demonstrator instrument has been used to demonstrate
more wavefront reconstruction algorithms on-sky than any previous \ao
system, including Learn and Apply \citep{2010JOSAA..27A.253V}, CuRe
\citep{cure}, Hierarchical Wavefront Reconstruction \citep{dicure},
Neural Networks \citep{ngslgstomoshort} and full \lqg control \citep{lqgshort}.
Because \canary is a visitor instrument, there are long periods of
time when it is either in storage or transport, or undergoing
laboratory integration.  The small number of on-sky nights each year
are therefore not ideal for testing new real-time implementations of
algorithms: These should be verified before reaching the telescope.
For this purpose, we have a real-time simulation code which is used
with the \canary \rtcs (which can run on a standard PC), allowing us
to verify these algorithms.

The necessity of this hardware-in-the-loop simulation was recently
demonstrated during development of the DiCuRe \scao reconstruction
algorithm \citep{dicure}.  This was first demonstrated on-sky in 2012.
Minor improvements were made off-line, and a few months later the
algorithm retried on-sky.  However, this time, performance was
degraded, and a spurious tilt signal was seen to develop with time on
the \dm.  The real-time simulation code was then used along with the
\rtcs in the following months to trace the source of this problem,
which was eventually found to be related to production of an actuator
mapping matrix.  Since this bug was not present in the non-real-time
code implementation used for development, it would not have been
possible to trace without the real-time simulation capability.

\subsection{Correlation wavefront sensing investigations}
The elongation of \lgs spots can be problematic for wavefront sensing,
with the \lgs signal spread over a larger number of pixels resulting
in lower signal-to-noise ratios and a reduction in sensitivity to
wavefront slope in the elongation direction.  Correlation based
wavefront slope estimation can be used to improve performance
\citep{2006MNRAS.371..323T,basden14}, though with additional complications.

While developing a correlation module for \canary, the use of the
real-time simulation facility was necessary to allow testing of
the real-time algorithm implementations (which differed significantly
from developmental versions), and to verify integration with the rest
of the \canary system, including coincident update of correlation and
slope references.  

We note that the simulation used here was not a hardware-in-the-loop
simulation in the strictest sense.  Our simulation did not operate at
real-time rates (a factor of three slower), and required
small changes to the \rtcs configuration, i.e.\ it was not hardware
anonymous:  The simulation interfaced to the \rtcs using Ethernet
sockets rather than the \sFPDP interface used by the \wfs cameras.

\subsection{CANARY}
A true real-time hardware-in-the-loop simulation for \canary will
greatly speed up algorithm development and improve the robustness of
the \canary software infrastructure.  This will allow new tomography
algorithms to be tested in their on-sky format and allow the interaction
between all components of \canary to be tested even when the equipment
is not available.  

A number of tomographic wavefront reconstruction algorithms used with
\canary require recorded data over a long time period (of
order minutes) to perform the necessary calibration procedures.
Without a real-time simulation capability, this is difficult to
achieve without significant effort spent developing separate offline
simulations and then converting between the simulation format and the
\rtcs format which can be an error prone process.  A full real-time
simulation facility for \canary would therefore ease this process.

An integrated simulation will also allow facilities such as telescope
offloading, and $C_n^2$ profiling to be performed using the standard
\canary tools, enabling further verification.

\subsection{Integration of adaptive and active optics systems}
Historically, \ao systems have been developed in isolation from the
telescope environment.  For \elt systems this will not be possible due
to the prevalence of active optics, and indeed, \ao components being
integrated with the telescope structure.  A real-time \ao simulation
facility is therefore essential for testing the interaction of \ao
with active optics, particularly when vibration control algorithms
such as \lqg \citep{lqgshort} are used.  Testing of the \ao interface
with telescope guiding systems also requires such a facility.

\subsection{Additional benefits}
In addition to enabling the laboratory integration of \elt \ao
systems, a real-time simulation capability also brings the benefit of
greatly increased simulation frame-rates, allowing a faster coverage in
the investigation of simulation parameter space while designing an \ao
system.  Current \ao simulations typically take many hours to model a
few seconds of telescope time, thus restricting the parameter space
that can be practically explored.  A real-time simulation would thus
enable a greater parameter space to be explored, allowing \ao system
designs to be further optimised.

\ignore{sFPDP to ethernet.}

\ignore{
\subsection{E-ELT integration}
As mentioned in previous sections, the \elts have a significant 
}

\section{Conclusion}
We have introduced the concept of hardware-in-the-loop simulation for
astronomical \ao systems, using the idea of a real-time simulation
capability.  We have shown that this capability will be essential to
enable the integration, validation and verification of \elt \ao
systems.  Although challenging, we have shown that achieving real-time
rates in simulation is possible using current processing
technology.  We have considered different scenarios for the
replacement of different physical hardware components with modelled
hardware, detailing the approaches that would be required in each
case.  Finally we have demonstrated a current need for this
hardware-in-the-loop simulation capability on the existing \canary \ao
system.  

\section*{Acknowledgements}
This work is funded by the UK Science and Technology Facilities
Council, grant ST/I002871/1.

\bibliographystyle{mn2e}
\bibliography{mybib}

\begin{thebibliography}{}

\bibitem[\protect\citeauthoryear{{Ass{\'e}mat}, {Wilson} \&
  {Gendron}}{{Ass{\'e}mat} et~al.}{2006}]{assemat}
{Ass{\'e}mat} F.,  {Wilson} R.,    {Gendron} E.,  2006, Opt.\ Express, 14, 988

\bibitem[\protect\citeauthoryear{{Babcock}}{{Babcock}}{1953}]{adaptiveoptics}
{Babcock} H.~W.,  1953, \pasp, 65, 229

\bibitem[\protect\citeauthoryear{{Basden}, {Geng}, {Myers} \&
  {Younger}}{{Basden} et~al.}{2010}]{basden9}
{Basden} A.,  {Geng} D.,  {Myers} R.,    {Younger} E.,  2010, Appl.\ Optics,
  49, 6354

\bibitem[\protect\citeauthoryear{Basden}{Basden}{2014}]{basden14}
Basden A.~G.,  2014, In press

\bibitem[\protect\citeauthoryear{{Basden}, {Ass{\'e}mat}, {Butterley}, {Geng},
  {Saunter} \& {Wilson}}{{Basden} et~al.}{2005}]{basden4}
{Basden} A.~G.,  {Ass{\'e}mat} F.,  {Butterley} T.,  {Geng} D.,  {Saunter}
  C.~D.,    {Wilson} R.~W.,  2005, \mnras, 364, 1413

\bibitem[\protect\citeauthoryear{Basden, Bharmal, Myers, Morris \&
  Morris}{Basden et~al.}{2013}]{basden12}
Basden A.~G.,  Bharmal N.~A.,  Myers R.~M.,  Morris S.~L.,    Morris T.~J.,
  2013, \mnras, 435, 992

\bibitem[\protect\citeauthoryear{{Basden}, {Butterley}, {Myers} \&
  {Wilson}}{{Basden} et~al.}{2007}]{basden5}
{Basden} A.~G.,  {Butterley} T.,  {Myers} R.~M.,    {Wilson} R.~W.,  2007,
  Appl.\ Optics, 46, 1089

\bibitem[\protect\citeauthoryear{{Basden} \& {Myers}}{{Basden} \&
  {Myers}}{2012}]{basden11}
{Basden} A.~G.,  {Myers} R.~M.,  2012, \mnras, 424, 1483

\bibitem[\protect\citeauthoryear{{Basden}, {Myers} \& {Gendron}}{{Basden}
  et~al.}{2012}]{basden10}
{Basden} A.~G.,  {Myers} R.~M.,    {Gendron} E.,  2012, \mnras, 419, 1628

\bibitem[\protect\citeauthoryear{{Bitenc}, {Rosensteiner}, Bharmal, Basden,
  Morris, Obereder, Dipper, Gendron, Vidal, Rousset, Gratadour, Martin, Hubert
  \& Myers}{{Bitenc} et~al.}{2013}]{dicure}
{Bitenc} U.,  {Rosensteiner} R.,  Bharmal N.~A.,  Basden A.,  Morris T.~J.,
  Obereder A.,  Dipper N.~A.,  Gendron E.,  Vidal F.,  Rousset G.,  Gratadour
  D.,  Martin O.,  Hubert Z.,    Myers R.,  2013, in Proc.\ Conf.\ Adaptative
  Optics for Extremely Large Telescopes 3 Tests of novel wavefront
  reconstructors on sky with canary

\bibitem[\protect\citeauthoryear{{Costille} \& {Fusco}}{{Costille} \&
  {Fusco}}{2012}]{2012SPIE.8447E..57C}
{Costille} A.,  {Fusco} T.,  2012, in Society of Photo-Optical Instrumentation
  Engineers (SPIE) Conference Series Vol.~8447 of Society of Photo-Optical
  Instrumentation Engineers (SPIE) Conference Series, {Impact of Cn$^{2}$
  profile on tomographic reconstruction performance: application to E-ELT wide
  field AO systems}

\bibitem[\protect\citeauthoryear{{Evans}, {Lehnert}, {Cuby}, {Morris},
  {Swinbank}, {Taylor}, {Alexander}, {Lorente}, {Clenet} \& {Paumard}}{{Evans}
  et~al.}{2008}]{eagleScience}
{Evans} C.~J.,  {Lehnert} M.~D.,  {Cuby} J.~G.,  {Morris} S.~L.,  {Swinbank}
  A.~M.,  {Taylor} W.~D.,  {Alexander} D.~M.,  {Lorente} N.~P.~F.,  {Clenet}
  Y.,    {Paumard} T.,  2008, in Adaptive Optical Components II. Edited by
  Holly, Sandor ; James, Lawrence. Proceedings of SPIE, Volume 141, pp. 120-124
  Vol.~7014 of Presented at the Society of Photo-Optical Instrumentation
  Engineers (SPIE) Conference, {Science Requirements for EAGLE for the E-ELT}.
pp~1--2

\bibitem[\protect\citeauthoryear{{Foppiani}, {Diolaiti}, {Baruffolo},
  {Biliotti}, {Bregoli}, {Cosentino}, {Delabre}, {Lombini}, {Marchetti},
  {Rossettini}, {Schreiber}, {Tomelleri}, {Conan}, {D'Odorico} \&
  {Hubin}}{{Foppiani} et~al.}{2010}]{2010SPIE.7736E..99F}
{Foppiani} I.,  {Diolaiti} E.,  {Baruffolo} A.,  {Biliotti} V.,  {Bregoli} G.,
  {Cosentino} G.,  {Delabre} B.,  {Lombini} M.,  {Marchetti} E.,  {Rossettini}
  P.,  {Schreiber} L.,  {Tomelleri} R.,  {Conan} J.-M.,  {D'Odorico} S.,
  {Hubin} N.,  2010, in Society of Photo-Optical Instrumentation Engineers
  (SPIE) Conference Series Vol.~7736 of Society of Photo-Optical
  Instrumentation Engineers (SPIE) Conference Series, {System overview of the
  Multi conjugated Adaptive Optics RelaY for the E-ELT}

\bibitem[\protect\citeauthoryear{{Fried}}{{Fried}}{1966}]{fried}
{Fried} D.~L.,  1966, Journal of the Optical Society of America (1917-1983),
  56, 1372

\bibitem[\protect\citeauthoryear{{Fusco}, {Cl{\'e}net}, {Meimon}, {Cohen},
  {Paufique}, {Petit}, {Gratadour}, {Michau}, {Amans}, {Dournaux}, {Jagourel},
  {Schnetler}, {Conan}, {Robert}, {Gendron}, {Rousset} \& {Hubin}}{{Fusco}
  et~al.}{2010}]{2010aoel.confE2002F}
{Fusco} T.,  {Cl{\'e}net} Y.,  {Meimon} S.,  {Cohen} M.,  {Paufique} J.,
  {Petit} C.,  {Gratadour} D.,  {Michau} V.,  {Amans} J.-P.,  {Dournaux} J.-L.,
   {Jagourel} P.,  {Schnetler} H.,  {Conan} J.-M.,  {Robert} C.,  {Gendron} E.,
   {Rousset} G.,    {Hubin} N.,  2010, in Adaptative Optics for Extremely Large
  Telescopes {ATLAS: the Laser Tomographic Adaptive Optics module for the
  E-ELT}

\bibitem[\protect\citeauthoryear{{Gendron}, {Vidal}, {Brangier}, {Morris},
  {Hubert}, {Basden}, {Rousset} \& {Myers}}{{Gendron}
  et~al.}{2011}]{canaryresultsshort}
{Gendron} E.,  {Vidal} F.,  {Brangier} M.,  {Morris} T.,  {Hubert} Z.,
  {Basden} A.,  {Rousset} G.,    {Myers} R.,  2011, \aap, 529, L2

\bibitem[\protect\citeauthoryear{{Gratadour}, {Sevin}, {Perret} \&
  {Brule}}{{Gratadour} et~al.}{2013}]{gratadour}
{Gratadour} D.,  {Sevin} A.,  {Perret} D.,    {Brule} J.,  2013, in Proc.\
  Conf.\ Adaptative Optics for Extremely Large Telescopes 3 Building a
  reliable, scalable and affordable rtc for ao instruments on elts.
p.~1

\bibitem[\protect\citeauthoryear{{Lardi{\`e}re}, {Conan}, {Bradley}, {Herriot}
  \& {Jackson}}{{Lardi{\`e}re} et~al.}{2008}]{2008SPIE.7015E.129L}
{Lardi{\`e}re} O.,  {Conan} R.,  {Bradley} C.,  {Herriot} G.,    {Jackson} K.,
  2008, in Society of Photo-Optical Instrumentation Engineers (SPIE) Conference
  Series Vol.~7015 of Society of Photo-Optical Instrumentation Engineers (SPIE)
  Conference Series, {Laser-guide-star wavefront sensing for TMT: experimental
  results of the matched filtering}

\bibitem[\protect\citeauthoryear{Morris, Gendron, Basden, Martin, Osborn,
  Henry, Hubert \& Sivo}{Morris et~al.}{2013}]{ngslgstomoshort}
Morris T.,  Gendron E.,  Basden A.~G.,  Martin O.,  Osborn J.,  Henry D.,
  Hubert Z.,    Sivo G.,  2013, in Proc.\ Conf.\ Adaptive Optics for Extremely
  Large Telescopes 3 "multiple object adaptive optics: Mixed ngs/lgs
  tomography"

\bibitem[\protect\citeauthoryear{{Reeves}, {Myers}, {Morris}, {Basden},
  {Bharmal}, {Rolt}, {Bramall}, {Dipper} \& {Younger}}{{Reeves}
  et~al.}{2012}]{dragon}
{Reeves} A.~P.,  {Myers} R.~M.,  {Morris} T.~J.,  {Basden} A.~G.,  {Bharmal}
  N.~A.,  {Rolt} S.,  {Bramall} D.~G.,  {Dipper} N.~A.,    {Younger} E.~J.,
  2012, in Society of Photo-Optical Instrumentation Engineers (SPIE) Conference
  Series Vol.~8447 of Society of Photo-Optical Instrumentation Engineers (SPIE)
  Conference Series, {DRAGON: a wide-field multipurpose real time adaptive
  optics test bench}

\bibitem[\protect\citeauthoryear{{Rosensteiner}}{{Rosensteiner}}{2011}]{cure}
{Rosensteiner} M.,  2011, \josaa, 28, 2132

\bibitem[\protect\citeauthoryear{Sivo, Kulcsar, Conan, Raynaud, Gendron,
  Basden, Vidal \& Morris}{Sivo et~al.}{2013}]{lqgshort}
Sivo G.,  Kulcsar C.,  Conan J.,  Raynaud H.,  Gendron E.,  Basden A.,  Vidal
  F.,    Morris T.~a.,  2013, Opt.\ Express

\bibitem[\protect\citeauthoryear{{Spyromilio}, {Comer{\'o}n}, {D'Odorico},
  {Kissler-Patig} \& {Gilmozzi}}{{Spyromilio} et~al.}{2008}]{eelt}
{Spyromilio} J.,  {Comer{\'o}n} F.,  {D'Odorico} S.,  {Kissler-Patig} M.,
  {Gilmozzi} R.,  2008, The Messenger, 133, 2

\bibitem[\protect\citeauthoryear{{Thomas}, {Fusco}, {Tokovinin}, {Nicolle},
  {Michau} \& {Rousset}}{{Thomas} et~al.}{2006}]{2006MNRAS.371..323T}
{Thomas} S.,  {Fusco} T.,  {Tokovinin} A.,  {Nicolle} M.,  {Michau} V.,
  {Rousset} G.,  2006, \mnras, 371, 323

\bibitem[\protect\citeauthoryear{{Vidal}, {Gendron} \& {Rousset}}{{Vidal}
  et~al.}{2010}]{2010JOSAA..27A.253V}
{Vidal} F.,  {Gendron} E.,    {Rousset} G.,  2010, \josaa, 27, 253

\bibitem[\protect\citeauthoryear{{von Karman}}{{von Karman}}{1948}]{vonkarman}
{von Karman} T.,  1948, Proceedings of the National Academy of Science, 34, 530

\bibitem[\protect\citeauthoryear{{Wang} \& {Ellerbroek}}{{Wang} \&
  {Ellerbroek}}{2012}]{2012SPIE.8447E..23W}
{Wang} L.,  {Ellerbroek} B.,  2012, in Society of Photo-Optical Instrumentation
  Engineers (SPIE) Conference Series Vol.~8447 of Society of Photo-Optical
  Instrumentation Engineers (SPIE) Conference Series, {Computer simulations and
  real-time control of ELT AO systems using graphical processing units}

\end{thebibliography}
\bsp

\end{document}